\documentclass[aps,prb,amsmath,amssymb,twocolumn,superscriptaddress,showpacs,longbibliography]{revtex4-1}
\usepackage{amsmath,amssymb}
\usepackage{graphicx}
\usepackage{appendix}
\usepackage[usenames,dvipsnames]{color}
\usepackage[colorlinks=true, citecolor=blue, linkcolor=blue, urlcolor=blue]{hyperref}
\usepackage{bm}

\usepackage[nottoc,numbib]{tocbibind}

\begin{document}

\title{Floquet-engineered light-cone spreading of correlations in a driven quantum chain}

\author{Mona H.~Kalthoff}
\email{mona.kalthoff@mpsd.mpg.de}
\affiliation 
{Max Planck Institute for the Structure and Dynamics of Matter, Luruper Chaussee 149, 22761 Hamburg, Germany}

\author{Dante M.~Kennes}
\affiliation 
{Dahlem Center for Complex Quantum Systems and Fachbereich Physik, Freie Universit\"at Berlin, 14195 Berlin, Germany}

\author{Michael A.~Sentef}
\email{michael.sentef@mpsd.mpg.de}
\affiliation 
{Max Planck Institute for the Structure and Dynamics of Matter, Luruper Chaussee 149, 22761 Hamburg, Germany}

\date{\today}

\begin{abstract}
We investigate the light-cone-like spread of electronic correlations in a laser-driven quantum chain. Using the time-dependent density matrix renormalization group, we show that high-frequency driving leads to a Floquet-engineered spread velocity that determines the enhancement of density-density correlations when the ratio of potential and kinetic energies is effectively increased both by either a continuous or a pulsed drive. For large times we numerically show the existence of a Floquet steady state at not too long distances on the lattice with minimal heating. Intriguingly, we find a discontinuity of dynamically scaled correlations at the edge of the light cone, akin to the discontinuity known to exist for quantum quenches in Luttinger liquids. Our work demonstrates the potential of pump-probe experiments for investigating light-induced correlations in low-dimensional materials and puts quantitative speed limits on the manipulation of long-ranged correlations through Floquet engineering. 
\end{abstract}
\maketitle


\section{Introduction}
The control of materials properties with light is a growing research field.\cite{basov_towards_2017} Theoretical proposals for using light to change properties of interacting-electron systems range from spin systems \cite{mentink_ultrafast_2015,eckstein_designing_2017,claassen_dynamical_2017,oka_floquet_2019} via one-dimensional Luttinger liquids and charge-density waves \cite{kennes_floquet_2018,Periodic_FRG_Kennes_2018} and nonequilibrium superconductors
\cite{raines_enhancement_2015,hoppner_redistribution_2015,peronaci_transient_2015,sentef_theory_2016,knap_dynamical_2016,kennes_transient_2017,sentef_light-enhanced_2017,murakami_nonequilibrium_2017,wang_light-enhanced_2018,kennes_light-induced_2018,dasari_transient_2018,hart_steady-state_2018,sheikhan_dynamically_2019,claassen_universal_2019} to correlated insulators.\cite{wang_producing_2017,tancogne-dejean_ultrafast_2018,walldorf_antiferromagnetic_2018,topp_all-optical_2018,peronaci_resonant_2018,golez_multi-band_2019} However, in practice one often has to deal with heating effects that can blur Floquet-engineered properties.\cite{mciver_light-induced_2018,sato_microscopic_2019}

A sweet spot in laser-driven correlated systems was identified in Ref.~\onlinecite{kennes_floquet_2018}, where it was shown that high-frequency driving avoids runaway heating and tunes the system across a phase transition. Similarly, resonant laser excitation with phonons \cite{forst_nonlinear_2011} has been demonstrated experimentally to lead to light-induced phases with enhanced interactions and induced order parameters.\cite{fausti_light-induced_2011,mitrano_possible_2016,pomarico_enhanced_2017} However, in ultrafast materials science it is difficult to assess the actual correlation lengths involved in the build-up of correlations in real time. This is drastically different for instance in cold atoms in optical lattices, where light-cone-like spreading of correlations was demonstrated .\cite{cheneau_light-cone-like_2012}

Here we show how light-cone spreading of correlations can also be triggered by high-frequency laser driving. By investigating a laser-driven one-dimensional quantum chain with real-time density matrix renormalization group (t-DMRG) calculations, we demonstrate how the spread of correlations is dictated by an instantaneous effective mode velocity that can be understood in terms of Floquet-renormalized effective Hamiltonian parameters. Moreover we investigate build-up of a Floquet steady-state for continuous laser driving and compare against thermal states. Finally we find that a sufficiently fast switch-on of the drive leads to a kink at the edge of the light cone, reminiscent of a dynamical phase transition after a quantum quench. Our combined results demonstrate the rich opportunities to tune correlations in periodically driven systems, provided that adequate off-resonant driving regimes can be identified. 


\section{Model and Method}
To analyze the influence of an electromagnetic driving field on a one-dimensional correlated chain of spinless fermions, we consider the Hamiltonian
    
    \begin{align}\nonumber
	\label{eq:Hamiltonian_fermionic}
	H\left(t\right)
	=
	&\sum_j\left[
	-\frac{J(t)}{2}
	c_{j}^\dagger  c_{j+1}+\mathrm{H.c.}\right.
	\\
	&+\left.U\left(n_{j}-\frac{1}{2}\right)\left(n_{j+1}-\frac{1}{2}\right)\right]\,.
	\end{align}
Here $U>0$ is the nearest-neighbor Coulomb interaction and $J(t)$ the hopping amplitude, which becomes time-dependent in the driven case (see below). The operator $c_{j}^{(\dagger)}$ annihilates (creates) a fermion on lattice site $j$, 
and $n_j = c_{j}^{\dagger}c_{j}$ is the local number operator. Throughout this paper we assume an infinite chain at half filling. The influence of a spatially uniform, time-dependent electric field is taken into account by performing the Peierls substitution~\cite{jauho_theory_1984}, yielding the time dependent hopping 
$J(t)=J_0 \,\mathrm{exp}\left[iA(t)\right]$, 
where $A(t)$ is the vector potential corresponding to an electric field $E(t)=-\partial_t A(t)$.	In the following we use $J_0=1$ as our unit of energy. 

\begin{figure}
	\centering
	\includegraphics[width= \columnwidth]{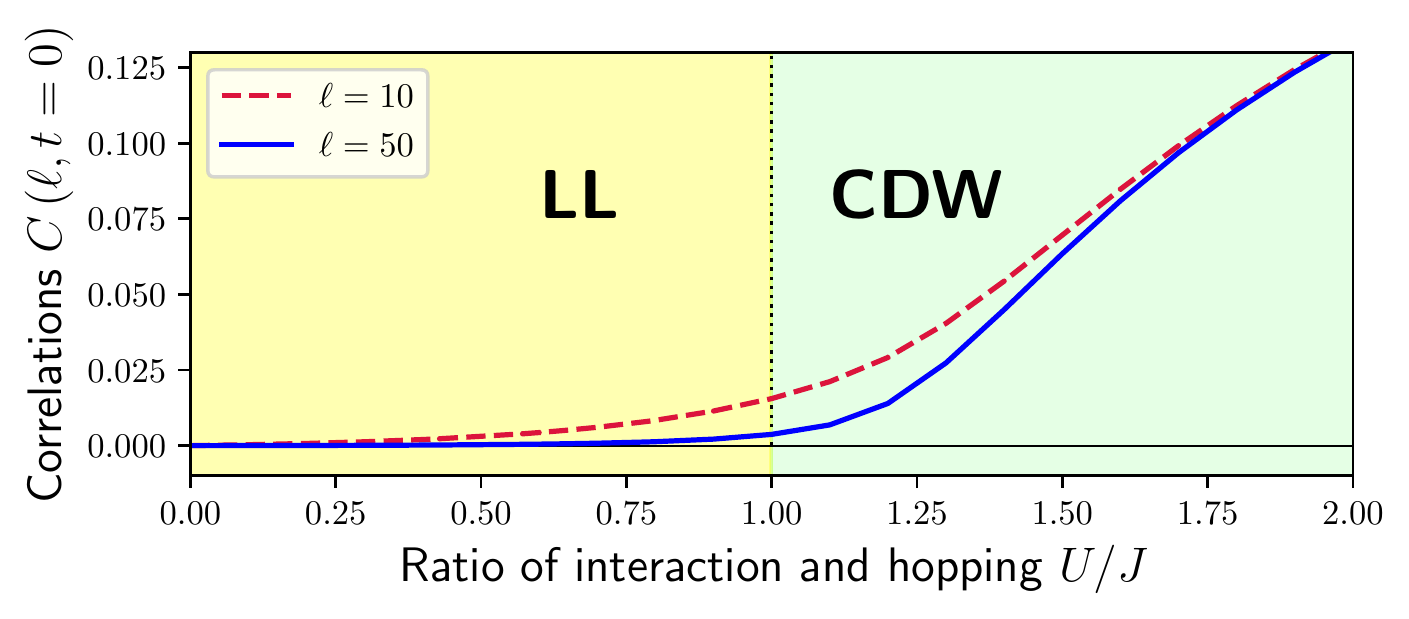}    
	\caption{Equilibrium quantum phase transition showing the transition from a Luttinger liquid (LL) to a charge-density wave (CDW) at $U/J=1$ with non-zero long-ranged correlations.
	}\label{fig:1Phasetransition}
\end{figure}

To set the stage for the nonequilibrium dynamics, we first characterize the well-known equilibrium phase diagram. The system has a quantum phase transition at $U/J=1$.
To characterize this phase transition, we compute the density-density correlation function
	\begin{align}
	\label{eq:Correlations}
    C\left(\ell,t\right)=\left\langle
    \left(n_0\left(t\right)-\frac{1}{2}\right)
    \left(n_\ell\left(t\right)-\frac{1}{2}\right)
    \right\rangle\,.
    \end{align}
This correlation function is shown for the groundstate ($t=0$) in Fig.~\ref{fig:1Phasetransition} for the 10$^\mathrm{th}$ (red) and the 50$^\mathrm{th}$ (blue) lattice site as a function of $U/J$. When $U/J < 1$ the system is a gapless Luttinger liquid (LL), which does not display long-range charge density wave order. For $U/J > 1$ a gapped charge density wave (CDW) phase emerges with staggered long-range density-density correlations. 

We now turn to the nonequilibrium dynamics. In the following we consider two different harmonic drives with frequency $\Omega$ in the high-frequency (Magnus) limit $\Omega \gg J,U$. The first is a drive that is ramped on over a time interval $\tau$, with vector potential 	
    \begin{align}
	A_\mathrm{D}(t) = \frac{E_0}{\Omega} \sin(\Omega t)\left[ 0.5+0.5 \mathrm{tanh}\left(\frac{t-5\tau}{\tau}\right)\right]\,.
	\label{eq:Driving_potential}
	\end{align}
This continuous-wave drive (CW-Drive) has previously been studied with the same DMRG method that we use in this paper.\cite{kennes_floquet_2018}
Hence, it is known that in the Magnus regime the growth of entanglement limiting the DMRG calculations remains manageable due to the absence of runaway heating, and the long-time limit is accessible. 

For high driving frequencies, it is known that a steady state can be defined in a parametrically long intermediate time regime.\cite{weidinger_floquet_2017,kennes_floquet_2018,claassen_dynamical_2017} Therefore, long-time physics can be described by a renormalized Hamiltonian, i.e., by a Hamiltonian that is averaged over a period of the drive. 
Averaging the time-dependent hopping $J(t)$ over the $2\pi/\Omega$ period yields the effective hopping 
    \begin{align}
	J_{\mathrm{eff}}=J_0\mathcal{J}_0\left(\frac{E_0}{\Omega}\right)\,,
	\label{eq:Jeff}
	\end{align}
where $\mathcal{J}_\alpha$ is the Bessel Function of the first kind.\cite{magnus_exponential_1954} Because the absolute value of the Bessel function is smaller than unity for any nonzero argument, the effective hopping is generically reduced compared to the equilibrium hopping. This implies that $U/J$ is increased by the drive. Therefore the laser drive moves the system to the right in the phase diagram shown in Fig~\ref{fig:1Phasetransition}, provided that the renormalization of $U/J$ is the dominant effect of the laser. Below we will show that this is indeed the case provided that the parameters of the problem are carefully chosen.  

Floquet theory~\cite{floquet_sur_1883} allows for an effective analytical study of periodically driven systems, but is restricted to time-periodic systems in analogy to Bloch theory for spatially periodic systems.
Therefore most theoretical studies of Floquet-driven systems \cite{shirley_solution_1965,sambe_steady_1973,oka_photovoltaic_2009,hemmerich_effective_2010,kitagawa_transport_2011,lindner_floquet_2011,rudner_anomalous_2013,iadecola_materials_2013,usaj_irradiated_2014,grushin_floquet_2014,dehghani_dissipative_2014,dehghani_out--equilibrium_2015,bukov_universal_2015,dalessio_dynamical_2015,eckardt_high-frequency_2015,mikami_brillouin-wigner_2016,eckardt_colloquium:_2017,claassen_dynamical_2017,uhrig_positivity_2019} assume the limit in which the drive was turned on in the infinitely distant past, which is impossible to realize within an experiment.
Nevertheless it has been shown that Floquet theory still captures the essential spectral features for system driven by laser pulses of finite duration~\cite{sentef_theory_2015,kalthoff_emergence_2018} if the system is probed on time scales sufficiently longer than the period of the driving field \cite{wang_observation_2013}, and if the pulse envelope is even longer than the probe duration. We therefore also consider a periodic drive that is modulated with a Gaussian envelope, given by 
	\begin{align}
	\label{eq:Gaus_potential}
	A_\mathrm{G}(t) = \frac{E_0}{\Omega} \sin(\Omega t)\mathrm{exp}\left[-\frac{(t-t_0)^2}{2\sigma^2}\right]\,,
	\end{align}
and compare our results to the system where the driving field is switched on and is kept switched on for long times.

\begin{figure*}
	\centering
	\includegraphics[width= \textwidth]{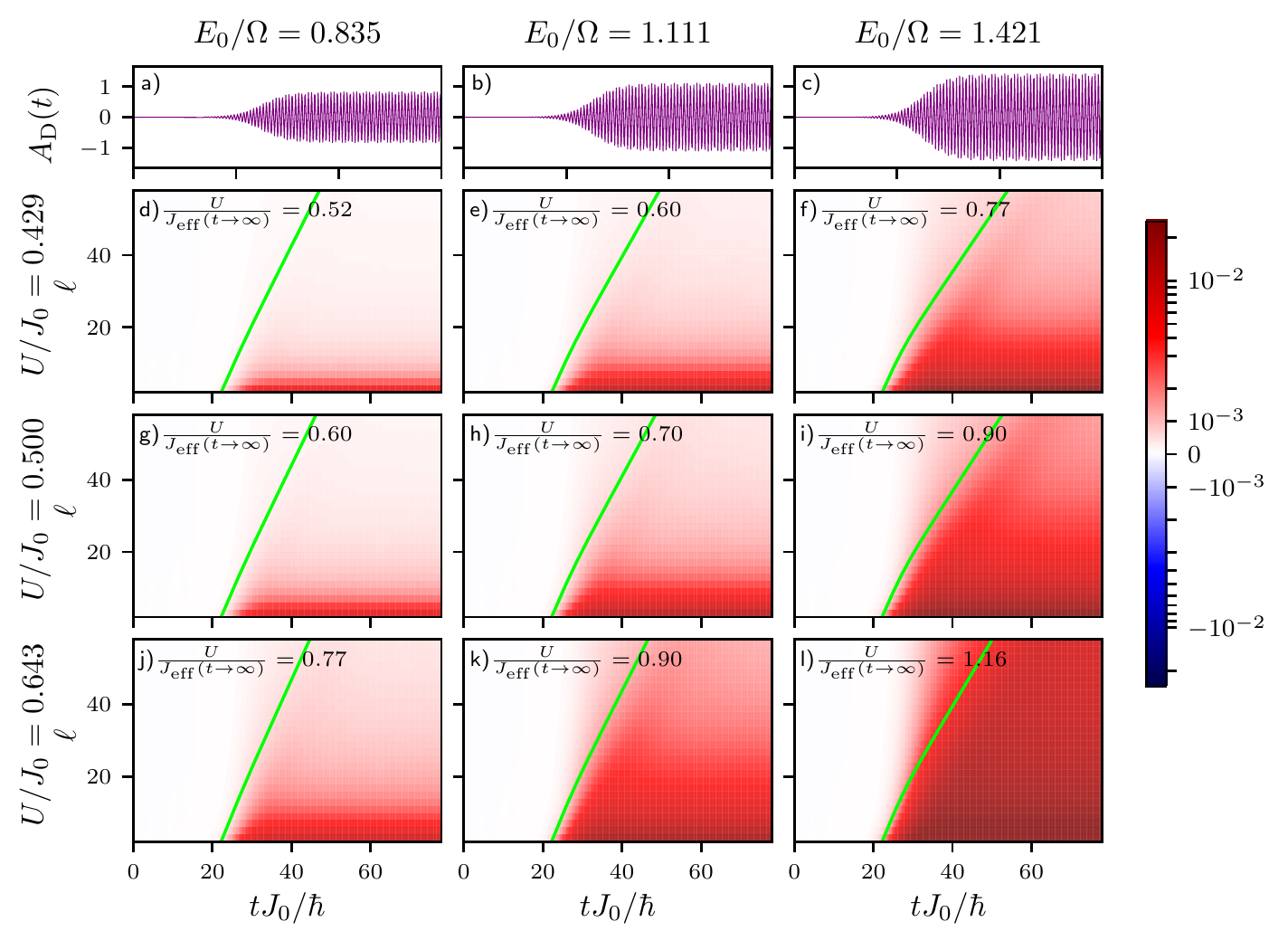} 
	\caption{{\bf Spread of correlations in the continuously driven chain.}
	(a)-(c) Vector potential $A_\mathrm{D}(t)$ of the ramped drives as a function of time with driving frequency $\Omega/J_0=15$ and ramp time $\tau J_0/\hbar=5$.
	(d)-(l) Heatmaps of correlation changes $C(\ell,t)-C(\ell,0)$ as a function of time $t$ for even distances $\ell$ and three different values of $U$ and $E_0/\Omega$, as indicated. $J_{\mathrm{eff}}\left(t_0\right)$ is the maximal amplitude of $A_\mathrm{D}(t)$.  
	}\label{fig:2Heatmaps_Driving}
\end{figure*}	

The spread of correlations within a quantum many-body system is restricted by a maximal velocity, known as the Lieb-Robinson bound.\cite{lieb_finite_1972} This bound is similar to the speed of light for the propagation of information in a relativistic quantum field theory. 
The corresponding light-cone effect has been demonstrated experimentally by quenching a one-dimensional quantum gas in an optical lattice.\cite{cheneau_light-cone-like_2012}
The spread of correlations can be visualized as modes departing from two lattice sites and the information being propagated when the modes interfere in the middle.\cite{lauchli_spreading_2008}
Therefore the velocity with which correlations spread through the lattice after it is excited by a quench is given by twice the maximal mode velocity.\cite{calabrese_time_2006}
In case of a LL, this means the expected velocity is given by
    \begin{align}
	2v_{\mathrm{LL}}=J\cdot\pi\cdot
	\frac{\sin(\arccos(-U/J))}{\arccos(-U/J)}\,.
	\label{eq:LL_velocity}
	\end{align}
To compare this velocity to our numerical data, where the system is not excited by a sudden quench, but a drive is turned on smoothly, we define a time-dependent effective hopping $J_{\mathrm{eff}}(t)$, which is calculated via the envelope functions of the drives. For the ramped case and the Gaussian pulse, this effective time-dependent hopping is given by
    \begin{subequations}
    \begin{align}
    J_{\mathrm{eff}}^{\mathrm{D}}(t)
	&=J_0
	\mathcal{J}_0
	\left[\frac{E_0}{\Omega}
	\left( 0.5+0.5 \mathrm{tanh}\left(\frac{t-5\tau}{\tau}\right)\right)
	\right],
	\\
	J_{\mathrm{eff}}^{\mathrm{G}}(t)
	&=J_0
	\mathcal{J}_0
	\left[\frac{E_0}{\Omega}
	\mathrm{exp}\left(
	-\frac{(t-t_0)^2}{2\sigma^2}\right)
	\right]\,,
	\label{eq:Jeff_timedependent}
	\end{align}
	\end{subequations}
respectively. Replacing $J$ in Eq.~\eqref{eq:LL_velocity} 
with $J_{\mathrm{eff}}(t)$ yields the spread velocity
    \begin{align}
	2v_{\mathrm{LL}}(t)=J_{\mathrm{eff}}(t)\cdot\pi\cdot
	\frac{\sin\left(\arccos\left(-U/J_{\mathrm{eff}}(t)\right)\right)}
	{\arccos\left(-U/J_{\mathrm{eff}}(t)\right)}\,.
	\label{eq:LL_velocity_time}
	\end{align}
Note that this implies that $J_{\mathrm{eff}}(t)$ decreases as the amplitude of the drive increases, and the velocity, which is dominated not by $U/J$ but by $J$, decreases because $J_{\mathrm{eff}}$ is smaller than one. Below we will check the validity of these expressions by comparing against the numerical data. 

\section{Results}
\begin{figure*}
	\centering
	\includegraphics[width= \textwidth]{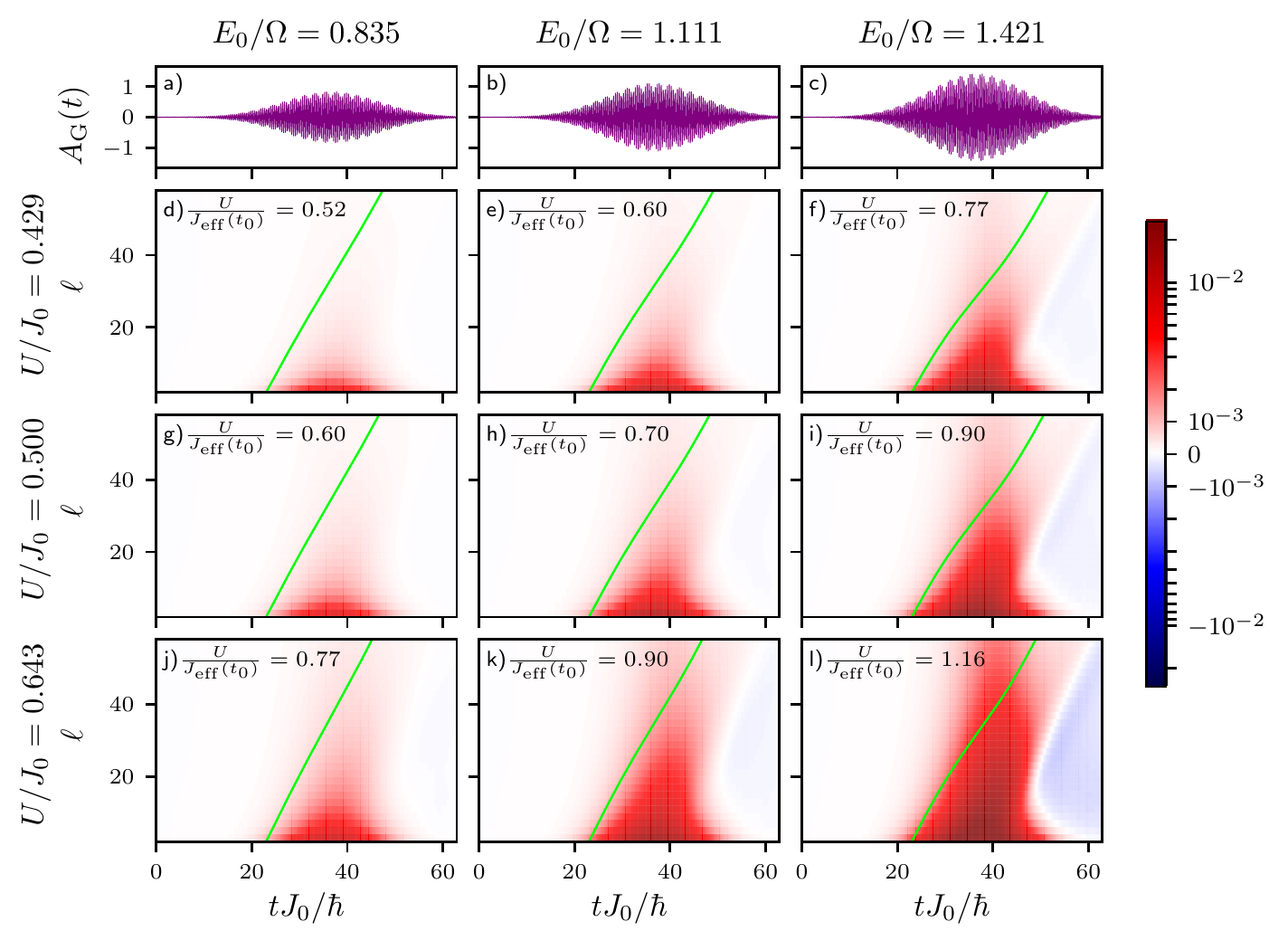}
	\caption{{\bf Spread of correlations in the pulse-driven chain.}
	(a)-(c) Vector potential $A_\mathrm{G}(t)$ of the Gaussian drives as a function of time with driving frequency $\Omega/J_0=15$, width $\sigma J_0/\hbar=10$, and $t_0 J_0/\hbar=36.77$.
	(d)-(l) Heatmaps of correlation changes $C(\ell,t)-C(\ell,0)$ as a function of time $t$ at even distances $\ell$ for three different values of $U$ and $E_0/\Omega$, as indicated.  
	}\label{fig:2Heatmaps_Gaus}
\end{figure*}

In this paper we consider three different interactions in the LL phase
$U/J_0=\left\lbrace 0.429,\,0.500,\,0.643\right\rbrace$ and three different maximal amplitudes $E_0/\Omega=\left\lbrace 0.835,\,1.111,\,1.421\right\rbrace$,
which we found to be representative. This yields values of $U/J_{\mathrm{eff}}$ between $0.52$ (LL) and $1.16$ (CDW) at the maximal driving amplitude. 
The corresponding driving field profiles are shown in the upper panels of Fig.~\ref{fig:2Heatmaps_Driving} and Fig.~\ref{fig:2Heatmaps_Gaus}, respectively. 
Note that the values for the interaction and the driving amplitudes are chosen such that even though $9$ different combinations of $U/J_0$ and $E_0/\Omega$ are given, there are only $6$ corresponding values of $U/J_{\mathrm{eff}}$. Thus the pairs (g) and (e), (f) and (j), as well as (i) and (k) have the same $U/J_{\mathrm{eff}}$.
The driving frequency is chosen to be $\Omega=15 J_0$, the ramp time of the ramped drive is $\tau J_0/\hbar=5$ unless denoted otherwise, and the parameters of the Gaussian pulse are given by 
$\sigma J_0/\hbar=10$ and $t_0 J_0/\hbar=36.77$.
The lower panels in Fig.~\ref{fig:2Heatmaps_Driving} and Fig.~\ref{fig:2Heatmaps_Gaus} display the heat maps of the light-induced changes of correlations at even distances $\ell$ (odd distances have opposite sign) as a function of time $t$. 
Here we subtract off the initial correlations, and $C(\ell,t)-C(\ell,0)$ is displayed on a logarithmic scale. 

The green lines show the expected spread of correlations, with twice the largest possible mode velocity in the LL, as defined in Eq.~\eqref{eq:Jeff},
integrated from $t_{\mathrm{start}}=22.2$ for the CW-Drive and $t_{\mathrm{start}}=23.0$ to $t$ for the pulse:
    \begin{align}
	\ell_{\mathrm{eff}}\left(t,t_{\mathrm{start}}\right)=
	\int_{t_{\mathrm{start}}}^t 2v_{\mathrm{LL}}(t')\,\mathrm{d}t'
	\label{eq:IntegralLL_velocity_time}
	\end{align}
Especially for small distances on the lattice, 
the wave front of the correlations computed with our t-DMRG data matches this green curve well, showing the validity of the concept of a time-dependent Floquet-engineered spread velocity. The magnitude of correlations increases as $U/J_{\mathrm{eff}}$ increases. 

At the maximal amplitude $J_{\mathrm{eff}}\left(t_0\right)$,
only the cases (d) to (k) in Fig.~\ref{fig:2Heatmaps_Driving} correspond entirely to the LL phase with ($U/J_{\mathrm{eff}}\left(t_0\right)=\left\lbrace 0.52,\,0.6,\,0.7,\,0.77\,0.9\right\rbrace$). 
The green curves match the data to a large extent in all of the shown cases, even at $U/J_{\mathrm{eff}}\left(t_0\right)=1.16$, which already reaches the CDW phase, even though Eq.~\eqref{eq:LL_velocity} should strictly only hold in the gapless LL phase. 
However, there are tails of correlations that exceed the $2v_\mathrm{LL}$ limit, which have a suppressed magnitude compared to the major wave front that spreads with $2v_\mathrm{LL}$ to a good approximation.

In order to compare more closely against time-resolved experiments, which typically employ a pump-probe setup with a pulsed driving envelope, we consider a sinusoidal field that is modulated with a Gaussian envelope. The correlation changes for such a Gaussian drive are displayed in Fig.~\ref{fig:2Heatmaps_Gaus}. Here the time-dependent effective velocity also matches the major wave front.
For the pulse we can also also observe the relaxation dynamics after the pulse. While $C(\ell,t)-C(\ell,0)$ is always non negative for the ramped drive, it does take slightly negative values when the field is switched off after the peak of the Gaussian pulse (see slightly blue areas in Fig.~\ref{fig:2Heatmaps_Gaus}). This effect can be understood as a consequence of heating (see discussion below and Appendix), with reduced correlations at effective nonzero temperatures compared to zero temperature. Moreover, we find that enhanced correlations last longer at larger distances, as can be seen from the red areas bending over to the right in Fig.~\ref{fig:2Heatmaps_Gaus}. This implies slower relaxation dynamics at longer distances, which is in accordance with the same light-cone effect that causes slower enhancement of correlations at longer distances when the drive is first switched on. 

\begin{figure}
	\centering
	\includegraphics[width=\columnwidth]{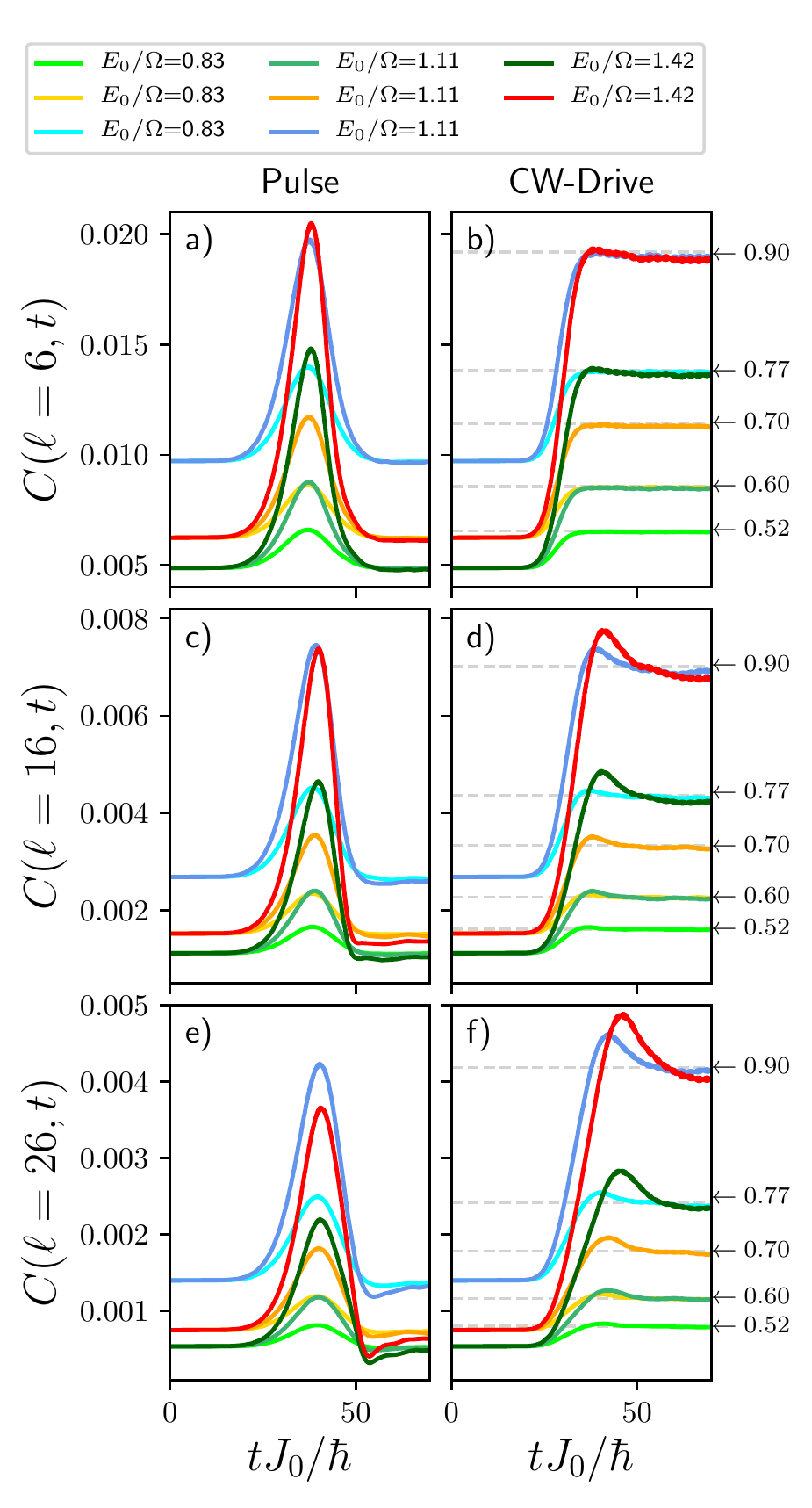}
	\caption{{\bf Temporal evolution of correlations at short, intermediate, and long distances.}
	Correlations $C(\ell,t)$ for a continuously driven system (left panels) and a pulsed system (right panels) with $\Omega/J_0=15$ and $\ell=6$ (top), $\ell=16$ (middle), and $\ell=26$ (bottom). Note the different scale on the $y$ axis for each $\ell$. Driving: $\tau J_0/\hbar=5$. Pulse: $\sigma J_0/\hbar=10$, $t_0 J_0/\hbar=36.77$. The grey lines indicate the correlations at thermal equilibrium (zero temperature) for the values of $U/J_{\mathrm{eff}}$ that are indicated with the arrows on the right.
	}\label{fig:End_of_chain}
\end{figure}

According to Floquet theory, a periodic drive with a driving frequency in the high-frequency Magnus regime should induce a Floquet steady state at sufficiently long times. In Fig.~\ref{fig:End_of_chain} the correlations for three different distances on the lattice are displayed as a function of time. The panels on the left ((a), (c), and (e)) display the correlations for a pulsed system, and the panels on the right ((b), (d), and (f)) display the correlations for the ramped drive. 
Note that the scale of the $y$ axis is different for all three distances (three rows), since the correlations are roughly ten times larger at $\ell=6$ than at $\ell=26$. 

For the ramped case, the correlations are stabilized to a steady state at short (b) and intermediate (d) distances, but as expected we find longer thermalization times for larges distances in the lattice (f). For the pulsed case, we find that the correlations at short (a) and intermediate (d) distances basically follow the driving pulse profile and return to the initial value almost perfectly shortly after the pulse, whereas for the longest distance shown here (e) the correlations are slightly suppressed below the initial value before thermalizing at longer times. 

The dashed grey lines in the panels on the right indicate the correlations at thermal equilibrium for the $U/J_{\mathrm{eff}}$ (whose value is indicated by the arrows on the right) corresponding to the curves. The magnitude of these equilibrium correlations does not depend on $U/J_0$ and $E_0/\Omega$ separately, but is solely determined by the value of $U/J_{\mathrm{eff}}\left(t\rightarrow\infty\right)$. Note that due to the fact that $3$ combinations of interaction strengths $U/J_0$ and maximal driving amplitudes $E_0/\Omega$ yield the same $U/J_{\mathrm{eff}}\left(t\rightarrow\infty\right)$, there are only $5$ different equilibrium correlations and corresponding steady-state values for the $8$ different cases shown. Indeed, we show in the Appendix that heating is negligible by computing the time-dependent energy absorption. The system evolves essentially adiabatically despite the fact that the LL phase is gapless and adiabaticity is not well-defined in this case. \cite{Adia_Dora_11,Adia_Kennes_17,Adia_Mazza_17}

\begin{figure}
	\centering
	\includegraphics[width= \columnwidth]{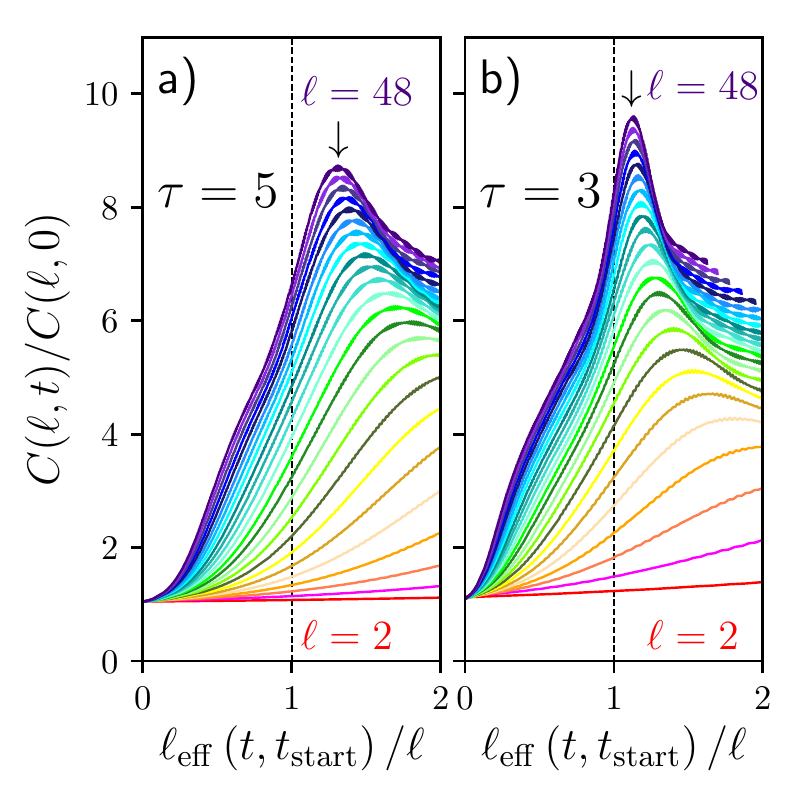}
	\caption{Renormalized correlations as a function of dimensionless distance at increasing distances $\ell$ in the lattice for ramp times $\tau J_0/\hbar=5$ (panel (a), same parameters as Fig~\ref{fig:2Heatmaps_Driving} (i)), and $\tau J_0/\hbar=3$ (panel (b)). Here $U/J_0=0.5$ and $U/J_{\text{eff}}(t \rightarrow \infty)=0.9$. Different curves correspond to increasing even lattice distances as indicated. 
	}\label{fig:Kink}
\end{figure}

We finally turn to the question of dynamical critical behavior when the system is driven between states with different correlation power laws. Within the LL phase, a scaling analysis shows that the correlations after a quench follow a
different power law inside compared to outside the light cone, which necessarily leads to a kink in the renormalized correlation function at the edge of the light cone. 
This can be shown analytically for quenched systems, for example in the interacting Tomonaga-Luttinger model.\cite{chudzinski_time_2016,collura_quantum_2015} In the following we identify an analogous kink at the edge of the light cone in our numerical data for driven systems.

To this end we show in Figure~\ref{fig:Kink} the correlation function  $C(\ell,t)/C(\ell,t=0)$, for two different switch-on times of the ramped drive, as a function of the effective dimensionless spreading distance, given by $\ell_{\text{eff}}\left(t, t_{\mathrm{start}}\right)/\ell$. The initial and final values of $U/J$ are within the LL phase in both cases. 
For large distances, where a power-law decay of correlations is expected, the kink should be located at $\ell_{\text{eff}}(t, t_{\mathrm{start}})/\ell$ $=$ $1$, i.e., at the edge of the light cone, which is indicated by the dashed line in Fig~\ref{fig:Kink}.

Indeed no clear peak can be identified at short distances in both cases, but a peak emerges and moves towards the edge of the cone at intermediate distances. At large distances the peak is well defined and approaches a kink-like discontinuity. As expected, the peak develops more clearly when the ramp time is shorter (Fig.~\ref{fig:Kink}(b)) compared to the slower ramp (Fig.~\ref{fig:Kink}(a)).
Interestingly this result proves that dynamical quantum criticality with nonanalytic behavior can indeed not only be observed for quantum quenches but also for Floquet-driven systems, paving the way for the potential observation of such critical behavior in pump-probe experiments on solids.

\section{Summary and Outlook}
In this work we have investigated in detail the Floquet-engineered spread of correlations in a driven quantum chain with Luttinger liquid and charge-density wave phases. In particular we have shown that light-cone effects exist even in driven systems with finite ramp times and finite laser pulses when the velocity renormalization due to the driving field is properly taken into account. 
Our findings prove that thermalization of correlations at moderate distances happens relatively quickly, provided that heating can be largely avoided in the first place. In our study heating is effectively suppressed although the Luttinger liquid is gapless and the switch-on time scale of the laser drive is relatively fast. The suppression of heating is enabled by the choice of an off-resonant, large driving frequency. 

The upshot from our results for laser-driven materials is that light-induced phase transitions and nonequilibrium materials engineering can be rationalized. In analogy with experiments on cold atomic gases \cite{cheneau_light-cone-like_2012} the effective correlation length that can be induced by nonequilibrium engineering of microscopic interactions is speed-limited only by the largest available relevant mode velocity. For example, for velocities on the order of $10^6$ ms$^{-1}$, which is a typical scale for the Fermi velocity in graphene, a correlation length on the order of $10^{-6}$ m is established within half a picosecond. Correspondingly, for slower modes like phonons, magnons, or plasmons, the times for correlations on the micrometer level are longer. It would be highly intriguing to devise experiments that are able to measure such time scales for the build-up of correlations in Floquet-engineered materials, which might be possible with time-resolved scattering at X-ray free-electron lasers.\cite{trigo_fourier-transform_2013,wall_ultrafast_2018,chen_theory_2019}

\textit{Acknowledgments.}
We acknowledge helpful discussions with Andrew Millis and Michael Sekania. M.H.K.~acknowledges financial support by the IMPRS-UFAST Program. Financial support by the DFG through the Emmy Noether program is gratefully acknowledged (D.M.K.: KA 3360/2-1; M.A.S.:SE 2558/2-1).

\bibliography{spin-chain-correlations}

\appendix
\section{Energy absorption}

\begin{figure}[htp!]
	\centering
	\includegraphics[width= \columnwidth]{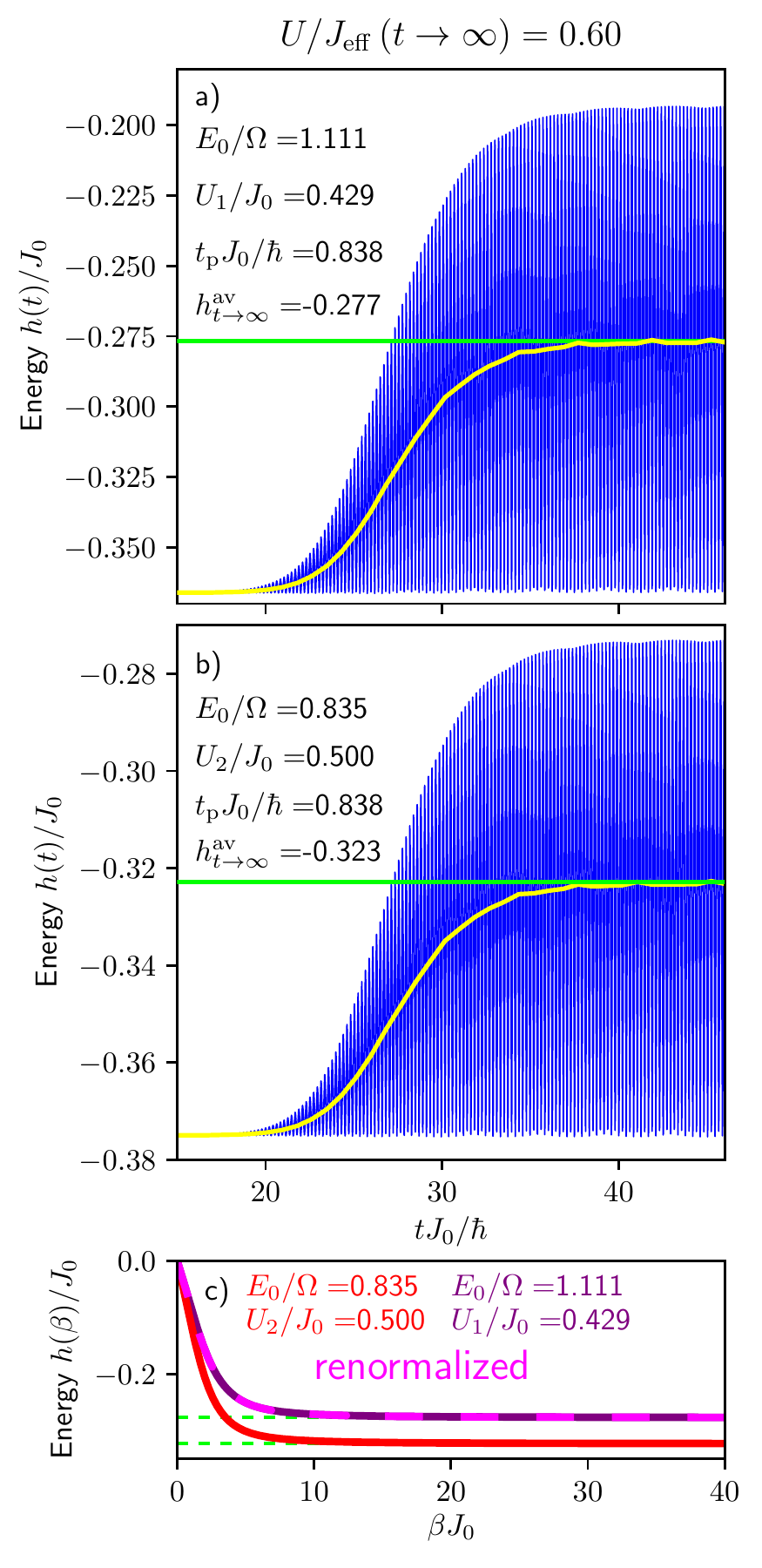}    
	\caption{(a), (b) Energy $h(t)$ as a function of time for $U_1/J_0=0.429$ and $E_0/J_0=1.111$ (a) and $U_2/J_0=0.500$ and $E_0/J_0=0.835$ (b), respectively, shown as rapdily oscillating blue curves. Averaging over the numerically determined period duration $t_{\mathrm{p}}J_0/\hbar$ leads to the period-averaged energy $h^\mathrm{av}\left(t\right)$ (yellow).   In addition, we indicate in the legend the long-time limit $h_{t\rightarrow\infty}^{\mathrm{av}}$. Green lines indicate the respective equilibrium energies of the Floquet-renormalized Hamiltonian at zero temperature. The frequency of the drive is $\Omega/J_0=15$ and the ramp time is $\tau J_0/\hbar=5$.
	(c) Energy $h(\beta)$ as a function of inverse temperature for $U_1/J_0=0.429$ and $E_0/J_0=1.111$ (purple, same parameters as Fig.~\ref{fig:2Heatmaps_Driving}(g)) and $U_2/J_0=0.500$ and $E_0/J_0=0.835$ (red, same parameters as Fig.~\ref{fig:2Heatmaps_Driving}(e)). A renormalization of $h(\beta)$ for the red curve ($U_1/J_0=0.500$, $E_0/J_0=0.835$) of both the $x$-axis and the $y$-axis with the ratio of energy scales, namely $U_1/U_2=0.858$, leads to the dashed magenta curve, which is identical to the purple curve ($U_1/J_0=0.429$ and $E_0/J_0=1.111$), showing that the curves are related by simple rescaling of all energies, as they should.}
	\label{fig:Energy}
\end{figure}

Here we briefly comment on the extraction of a Floquet steady-state energy in comparison with the groundstate energy of the Floquet-renormalized Hamiltonian, which proves that energy absorption is minimal and heating is avoided in the high-frequency driving regime. 
The upper two panels in Fig.~\ref{fig:Energy} display the Energy $h(t)$ as a function of time. The values of $U/J_0$ and $E_0/\Omega$ are chosen such that $U/J_{\mathrm{eff}}\left(t\rightarrow\infty\right)=0.60$ for both panels (analogous to (e) and (g) 
in Fig~\ref{fig:2Heatmaps_Driving}). The numerically extracted period duration of the time-dependent energy, $h^\mathrm{av}\left(t\right)=0.838$, equals twice the period of the drive. By comparison with the groundstate energy for the Floquet-renormalized parameters with $U/J_{\mathrm{eff}} = 0.60$, it becomes evident that the time-averaged energy of the driven system approaches exactly this groundstate energy in the long-time limit. This proves that heating is indeed minimal, as discussed in the main text in the context of Fig.~\ref{fig:End_of_chain}.

Finally we illustrate that the energies corresponding to parameters that have the same $U/J_{\mathrm{eff}}\left(t\rightarrow\infty\right)=0.60$ are directly related to each other. Fig.~\ref{fig:Energy}(c) displays the energy as a function of the inverse temperature $\beta$ and shows that they are simply related by rescaling of all energies in the problem for the two cases shown.

\end{document}